\documentclass[12pt]{iopart}
\usepackage{graphicx}

\begin{document}

\title{Interisotope effects in optimal dual-isotope loading into a shallow optical trap}
\author{M S Hamilton, A R Gorges and J L Roberts}
\address{Department of Physics, Colorado State University, Fort Collins, Colorado 80523, USA}
\ead{mathamtn@rams.colostate.edu}

\begin{abstract}
Examination of loading the isotopes $^{85}$Rb and $^{87}$Rb simultaneously into a shallow far-off-resonance trap (FORT) has revealed an unexpected decrease in maximum atom number loaded as compared to loading either isotope alone.  
The simultaneous loading of the FORT will be affected by additional homonuclear and heteronuclear light-assisted collisional losses.
However, these losses are measured and found to be insufficient to explain the observed drop in total number of atoms loaded into the FORT.  
We find that our observations are consistent with a decrease in loading rate caused by inter-isotope disruptions of the efficient laser cooling required to load atoms into the optical trap.
\end{abstract}


\submitto{\jpb}
\maketitle

Mixtures of multiple atomic species at ultracold temperatures exhibit interactions which are useful in many applications.
For example, multi-species systems allow for the formation of ultracold molecules~\cite{barletta:052707,hudson:063404,kohler:1311,Kraft2006,papp:180404,Schloder2001,Soldan2004}, formation of multi-species quantum-degenerate systems~\cite{catani:011603}, and cooling of one species by another via sypethetic cooling~\cite{Soldan2004,Anderlini2005,Bloch2001,Haas2007,Hadzibabic2002}.
Multi-species systems are also useful in ultracold chemistry~\cite{bell2009} and electron dipole moment~\cite{Chin2001} experiments.  
The applications in ultracold chemistry may also provide a novel means for quantum information processing~\cite{DeMille2002}.

A far-off resonant trap (FORT) is a useful tool for studying multi-species systems.
A FORT has the ability to hold atoms and molecules~\cite{Takekoshi1998,Jochim2003,Strecker2003,Zirbel2008,Ospelkaus2008} for long periods of time with minimal heating from rescattered photons~\cite{Miller1993}.
Any magnetic sublevel can be confined in a FORT~\cite{Takekoshi1998} which allows for the formation of quantum degenerate systems that would not be possible using magnetic fields~\cite{Barrett2001,Weber2003,Takasu2003,Cennini2003,Griesmaier2006,Dumke2006,Gericke2007,Gross2008,beaufils:061601,Granade2002,Fukuhara2007}.
However, there are limitations associated with trapping and loading more than one species at a time into a FORT.

Often FORTs are loaded from magnetic traps~\cite{Lin2009}.
However, it is more experimentally straightforward to load a FORT directly from a magneto-optical trap (MOT).
The maximum number of atoms loaded into the optical trap is an important consideration in almost any experiment.
This number is determined by the balance between the loading rate of atoms into the FORT from the MOT and light-assisted collisions that produce losses in the FORT~\cite{Kuppens2000}.
When loading two species into a FORT, each species to be loaded requires laser light at appropriate frequencies.
The need for additional trapping lasers can hinder the efficiency of loading multiple species into the FORT since their presence increases the number of loss channels as compared to loading a single species alone.
There are two sources of these additional loss channels: on- and off-resonant heteronuclear losses between isotopes, and off-resonant homonuclear losses.
It is not immediately clear how much of an impact the additional losses will have on dual-isotope loading.
Heteronuclear loss rates and off-resonant homonuclear loss rates are expected to be smaller than the on-resonant homonuclear loss rate because in general the collisions occur at shorter internuclear separation~\cite{Weiner1999}.
However, there are many additional loss channels.
The ultimate effect on the performance of the FORT will depend on the comparison of the single-isotope loss rates with the sum of the large number of additional smaller-rate loss channels.

We have optimized dual-isotope loading of $^{85}$Rb and $^{87}$Rb into a shallow FORT.
By taking the appropriate measurements, we can quantitatively measure the relevant loss rates.
The observed loading can then be quantitatively compared with the loading expected from a model.
This allows us to determine how much an impact the heteronuclear and off-resonant homonuclear losses have on the loading process under optimized conditions, as well as if there are any other mechanisms affecting load performance other than these additional losses.
We find that the sum of the heteronuclear loss rates is approximately the same as the sum of the single-isotope on-resonant homonuclear loss rates in our system.
The total off-resonant loss rates are only about half as large as the sum of all the heteronuclear loss rates.
Individual heteronuclear loss rates along an individual loss channel are much smaller than those associated with homonuclear losses, but they have many more possible loss channels.
Comparing our observed loading performance to a quantitative model, we find that these additional loss rates alone are insufficient to explain the observed reduction in the number of $^{85}$Rb and $^{87}$Rb trapped, however.
Instead, the difference between the observed maximum number of atoms trapped and the number predicted including the additional losses can be explained through reductions in the load rate caused by a drop in efficiency in laser cooling due to the presence of the other isotope.
This paper details our quantitative measurements of dual-isotope loading from a MOT to a FORT.

It is beneficial to begin our discussion with the case of single-isotope loading.  As previously mentioned, the number of atoms loaded into the FORT is a function of two competing processes: the rate of atoms loaded into the FORT and collisional losses.  
The rate of atoms loaded into the FORT is determined by the temperature of the atoms in the MOT, the number of atoms that enter the load volume per unit time, and the effectiveness of the cooling light in slowing the atoms in order that they become confined by the conservative optical potential.
Losses are primarily induced by light-assisted collisions, but may also include other losses such as collisions with background gas atoms.
Thus, the number of single-isotope atoms loaded into the FORT is described by the equation~\cite{Kuppens2000}
\begin{eqnarray}
\case{\rmd N}{\rmd t}=R(t) - \Gamma N - \beta '\case{N^{2}}{V},
\label{aloneEq}
\end{eqnarray}
where \textit{N} is the number of atoms in the FORT, \textit{V} is the volume of the trap, \textit{R}(\textit{t}) is the load rate of atoms into the trap, $\Gamma$ characterizes single-body losses due to collisions with background gas atoms, and $\beta^{\prime}$ is an effective two-body loss coefficient.  
On the time scales used in our experiment, single-body losses contribute much less than the two-body losses and thus the $\Gamma$ term may be approximated as zero in our model. 
There is an observable time dependence of the load rate \textit{R}(\textit{t}) which is caused by changes in our MOT density and position during the loading process~\cite{Hamilton2009}.
The effective loss coefficient depends on not only the intensity and detuning of the various lasers used during loading, but also the \textit{F} and $m_{F}$ population distributions of the trapped atoms.
Nevertheless we observe that the effective loss coefficient $\beta^{\prime}$  has little to no variation during the course of loading, and can therefore be treated as a constant.

The description of optical trap loading is very similar in the two-isotope case.
However, there are the additional two-body loss mechanisms discussed above.
Therefore, dealing with both isotopes requires two coupled differential equations of similar form as (\ref{aloneEq}), with the addition of a two-body cross-loss term:
\begin{eqnarray}
\eqalign{\fl\case{\rmd N_{85}}{\rmd t}=R_{85}(t,N_{87},N_{87 \rm{MOT}}) - \Gamma _{85}N_{85}\\
 - \beta _{85}'\case{N_{85}^{2}}{V} - \beta _{85-87}'\case{N_{85}N_{87}}{V}
\cr
\fl\case{\rmd N_{87}}{\rmd t}=R_{87}(t,N_{85},N_{85 \rm{MOT}}) - \Gamma _{87}N_{87}\\
 - \beta _{87}'\case{N_{87}^{2}}{V} - \beta _{85-87}'\case{N_{87}N_{85}}{V}.}
\label{dualEq}
\end{eqnarray}
In (\ref{dualEq}), subscripts have been added to explicitly denote each isotope of Rb. 
The effective two-body cross-loss coefficient $\beta _{85-87}^{\prime}$, assumes that cross-species collisions result in equal losses of both Rb isotopes.
This is reasonable given the mass of each isotope is close to the other and the FORT trap depth is the same for both isotopes.
Again, the single-body losses do not contribute significantly in our apparatus, so the terms $\Gamma _{85}$ and $\Gamma _{87}$ can be ignored.
In addition, it should also be noted that all $\beta^{\prime}$ terms in~(\ref{aloneEq}-\ref{dualEq}) are effective two-body loss coefficients, and thus represent a sum over individual loss channels.
The dependence of the load rate \textit{R} on the presence of the other isotope was initially unexpected.
This cross-isotope hindrance to the loading will be discussed in further detail below.
One of the ways this reduction in load rate manifests itself is through a decrease in the maximum number of atoms loaded into the FORT.

In order to examine performance of dual-isotope loading, the following experimental procedure was followed.
Two overlapping magneto-optical traps are prepared, using standard techniques~\cite{Raab1987}, inside a chamber containing a thermal Rb vapor.
Each MOT traps either $^{85}$Rb or $^{87}$Rb atoms and consists of its own hyperfine pump laser as well as a trapping light field.
The trapping light field is formed by a laser beam which is split and then retroreflected.
Each of the six trapping beams in the MOT has an average peak intensity of 2.5 mW/cm$^{2}$ for $^{85}$Rb and 4.8 mW/cm$^{2}$ for $^{87}$Rb.
The laser beams of either MOT can be allowed into the chamber using separate shutters, giving us the ability to take measurements with either isotope of Rb alone or both isotopes at the same time.
We allow each of the MOTs to accumulate 2$\times$10$^{8}$ atoms prior to starting the loading sequence.
The FORT is produced by a CO$_{2}$ laser operating at a wavelength of 10.6~$\mu$m, which is turned on and off non-adiabatically (less than 1~$\mu$s) via an acousto-optical modulator.
The FORT itself consists of a 30~W beam with a trap depth of 120~$\mu$K whose focus is overlapped with the MOT region.

A typical load of the FORT is accomplished using a series of compressed MOT (CMOT)~\cite{Petrich1994} configurations where the trapping laser detuning is further increased to the red of the cycling transition and the hyperfine pump power is significantly reduced.
The CMOT stages are followed by an optical molasses stage where the anti-Helmholtz coils are switched off and the trapping lasers are further detuned.
During the optical molasses stage, the $^{85}$Rb trapping laser was set to 80~MHz to the red of its cycling transition, and $^{87}$Rb laser was set to 120~MHz to the red of its cycling transition.
It is also during this stage that the FORT is turned on and loading occurs.
The duration of this stage is referred to as the FORT loading time and is adjusted to examine loading.
When we wish to stop loading, the hyperfine pump lasers are turned off for 1~ms prior to the MOT trapping lasers.  
This puts all the atoms into the \textit{F}=2 or \textit{F}=1 hyperfine state for $^{85}$Rb and $^{87}$Rb respectively.
Atom temperature after loading is about 15~$\mu$K.
The atoms are held in the FORT for 100~ms to allow any residual MOT atoms to fall away.
The FORT is then turned off and the atoms are allowed to expand for 5~ms prior to being imaged by a charged-coupled device camera using absorption imaging.
The resulting image is then analyzed to determine the atom number loaded in the FORT.

\begin{figure}
\includegraphics{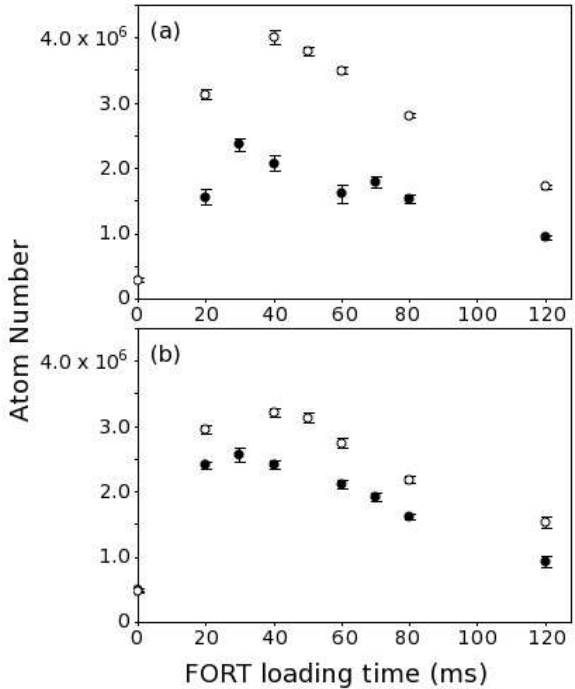}
\caption{\label{Evos} Typical evolution of atom number loaded into the FORT as a function of time for both (a) $^{85}$Rb and (b) $^{87}$Rb.  
Open circles denote atom number during loading of the isotope alone, while full circles denote the isotopic atom number during loading while loading with the other isotope.  
Error bars denote statistical uncertainties.}
\end{figure}

When both isotopes of Rb are loaded at the same time, we observe that the sum of the maximum number of atoms for each isotope trapped by the FORT drops by as much as a factor of 2 compared to what might be expected by summing the maximum number loaded of each isotope alone.
This is seen in figure~\ref{Evos}, which depicts the atom number loaded into the FORT as a function of load time for both alone (open circles) and dual (full circles) loading of (a) $^{85}$Rb and (b) $^{87}$Rb atoms.
Figure data for loading of an isotope alone was taken without the MOT light of the other isotope present.
With the MOT light of the other isotope present, a decrease in maximum atom number of almost 10\% in $^{85}$Rb and less than 3\% in $^{87}$Rb has been observed, which is less than the reduction depicted in figure~\ref{Evos}.
In order to determine the source of this reduction, we first compared the measured dual-isotope FORT loading behaviour to the behaviour predicted by the model described by (\ref{dualEq}), using measured loss rates and under the initial assumption that the load rate was independent of the number of isotopes being loaded.
To perform this comparison, we conducted a series of experiments designed to measure the load rate and all the coefficients $\beta^{\prime}$ under conditions relevant to the FORT loading.
In each of these experiments, the general strategy was to alter our normal loading conditions to make one of the terms dominant over the others so that the dominant term could be measured.
For example, to measure load rate the number of atoms in the optical trap was kept small by delaying the time at which it was turned on; this ensures that the rate term dominates in~(\ref{aloneEq}-\ref{dualEq}).
Once all of the load rates and loss coefficients were determined, the match between the model prediction and the measured data could be used to evaluate the validity of the model under the independent load rate assumption.

Under our initial independent load rate assumption, the load rate for either isotope was obtained from examination of single-isotope loading into the FORT.
This is significantly easier than extracting the load rate directly from dual-isotope loading data.
To solve for the coefficients in (\ref{aloneEq}), we examined the loading at the peak of the loading curve where dN/dt is equal to zero following the technique of~\cite{Hamilton2009}.
This corresponds to about 40~ms of loading for both sets of open point data depicted in figure~\ref{Evos}.
By delaying the turn on time of the FORT beam to that of the peak and taking short duration measurements, we obtained the load rate at the peak.
This was then used to solve for the constant loss coefficient $\beta^{\prime}$ using the full loading behaviour at the peak.
The loss coefficient for $^{85}$Rb and $^{87}$Rb was observed to be 6.3$\pm$0.7$\times$10$^{-11}$~cm$^{3}$/sec and 9.2$\pm$1.2$\times$10$^{-11}$~cm$^{3}$/sec, respectively.
Once we obtained the single-isotope loss coefficient and confirmed experimentally that it was constant in time, we then modeled the time dependence of the load rate using the load rate at the peak as a constraint.
A second order polynomial was sufficient to model the load rate over the time interval measured.

In order to accurately use the effective loss coefficients for dual-isotope loading, the hyperfine state distribution had to be measured.
This is because the loss rates are hyperfine state dependent~\cite{Gorges2008}.
Typically, during the imaging of atoms trapped in the FORT the hyperfine pump laser light is turned on to put atoms into the upper (\textit{F}=3 for $^{85}$Rb or \textit{F}=2 for $^{87}$Rb) hyperfine state and into resonance with the probe light.
To determine the hyperfine state distribution, we repeated the atom number measurements but rapidly shut off all other light followed by imaging without the hyperfine pump laser on.
This gave us the number of atoms in the upper hyperfine state which could then be compared with the full number of atoms from a standard measurement.
We found that during our dual loading conditions approximately ten percent of atoms are in the upper hyperfine state for $^{85}$Rb and about twenty percent for $^{87}$Rb.

The hyperfine state distribution was then used to calculate the homonuclear loss coefficients in combination with the homonuclear loss rates of individual state loss channels.
To extract values for these loss rates, we loaded the FORT with the isotope of interest to its peak number before turning off its trapping laser or hyperfine pump laser.
This will destroy that isotope's MOT, allowing those MOT atoms to fall away and cease the loading of that isotope into the FORT.
The atoms which were already loaded into the FORT remain, and are rapidly pumped into a single hyperfine state.
In the case of $^{85}$Rb, turning off the trapping laser will put the atoms into the \textit{F}=3 hyperfine state and turning off the hyperfine pump laser will put them into the \textit{F}=2 hyperfine state.
Similarly for $^{87}$Rb, shutting off the trapping laser puts the atoms into the \textit{F}=2 hyperfine state, while blocking the hyperfine pump laser puts the atoms into the \textit{F}=1 hyperfine state.
The decay of the atom number remaining in the FORT was then measured and then fitted to extract the hyperfine state-specific loss coefficient.

The homonuclear losses during single-isotope loading are not the same as homonuclear losses during dual-isotope loading.
This is because the addition of the second isotope's MOT lasers introduces off-resonant losses.
We explicitly measured these by examining the decay of one isotope loaded into the FORT alone with the off-resonant light of the other isotope turned on.
Loss rates driven by off-resonant light are reported in Table~\ref{crossloss}.
\Table{\label{crossloss}Measured hyperfine state-dependent losses of both isotopes of Rb in units of $10^{-11}$cm$^{3}/$sec.
Off-resonant single species losses are the homonuclear losses experienced by the isotope pumped into the indicated hyperfine state while subjected to the off-resonant trapping light of the other isotope.
Cross species losses refer to the losses between the indicated isotope and hyperfine state with the other isotope having a hyperfine state distribution typical of the distribution during a dual-isotope load.
The numbers in parenthesis indicate the statistical uncertainties for each measurement.
In addition, there is an overall systematic uncertainty of about 50~percent in these measurements.}
\br
Hyperfine State & Off-Resonant & \\
Dependent Loses & Single Species & Cross Species\\
\mr
$^{85}$Rb \textit{F}=3 & $\06.54(0.37)$ & $16.81(0.63)$ \\
$^{85}$Rb \textit{F}=2 & $\01.77(0.17)$ & $\04.33(0.47)$ \\
$^{87}$Rb \textit{F}=2 & $11.71(0.83)$ & $\08.07(0.40)$ \\
$^{87}$Rb \textit{F}=1 & $\00.36(0.21)$ & $\00.08(0.39)$ \\
\br
\endtab
The trends in loss rates depicted in Table~\ref{crossloss} are a result of various laser powers and detunings, as well as the complex inter-atomic potentials resulting from the hyperfine structure and distributions of the two isotopes.
We find that the off-resonant homonuclear losses of the upper hyperfine states are about an order of magnitude greater than without the off-resonant light present\footnote{Values for the on-resonant homonuclear losses quoted in the text are not directly comparable to the values listed in Table~\ref{crossloss}.
This is because the values quoted for the on-resonant losses are a mixture of states while the off-resonant losses are measured with a specific hyperfine distribution}.
The values in Table~\ref{crossloss} have an overall uncertainty of about 50\% due to systematic uncertainties in both number calibration and trap volume determination.
However, since these uncertainties are expected to apply equally to all measurements the relative comparisons have a precision reflected by the statistical uncertainty quoted in the table.

Once the homonuclear losses are known, the decay measurements were then repeated with both isotopes loaded into the FORT to give us the hyperfine state dependent heteronuclear losses.
This was done by modeling the decay of both isotopes present using the known homonuclear losses and fitting the heteronuclear losses to the data.
This method requires knowledge about the time dependent behaviour of the other isotope during the observed decay from the FORT.
We find that the atom number of the other isotope can be well modeled by a simple interpolating function.
This allows us to solve either part of (\ref{dualEq}), with the rate term set equal to zero, as a decoupled solitary differential equation.
All of the hyperfine state dependent losses are summarized in Table~\ref{crossloss}.
Upper state (F=3 for $^{85}$Rb and F=2 for $^{87}$Rb) heteronuclear losses are observed to be much larger than the lower state (F=2 for $^{85}$Rb and F=1 for $^{87}$Rb) heteronuclear losses.
The larger upper state losses are not surprising since the relatively strong trapping lasers are nearly in resonance with those states.

\begin{figure}
\includegraphics{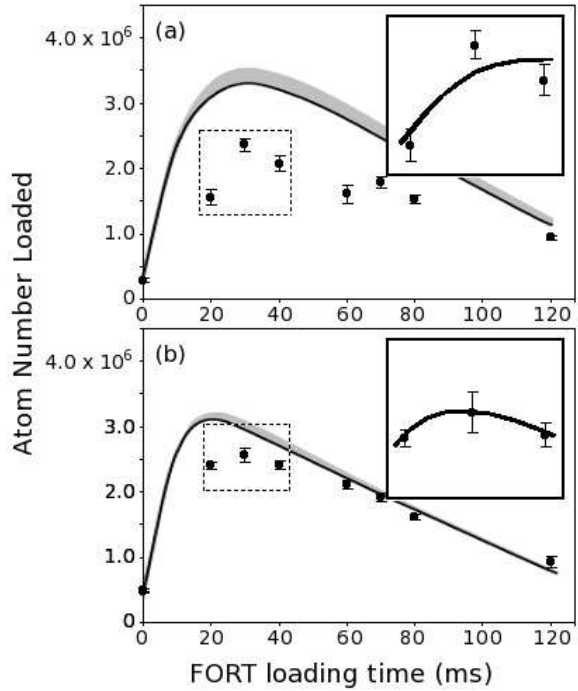}
\caption{\label{Dual} Model of dual-isotope evolution of atom number loaded into the FORT under the independent load rate assumption along with the actual dual loading data.
Plotted separately are the individual isotopes of (a) $^{85}$Rb and (b) $^{87}$Rb, with the total number of atoms in the trap being the sum of the two.
The points are the experimental values with error bars representative of statistical error of the measurement.
The curves follow the model behaviour of the coupled differential equations given in~(\ref{dualEq}) with the observed load rate from single isotope loading and losses calculated explicitly from measured rates from individual loss channels.
Due to our inability to separately determine individual channel loss rates (F=2+2 vs. F=2+1) that make up the 87 effective homonuclear loss rate $\beta^{\prime}_{87}$, the model prediction is shown as a band of possible values.
Our observations indicate that actual behaviour is likely to be close to the solid line.
The insets show the best fit allowing for the variation of the load rate due to the presence of the other isotope near the peaks of the loading curves.}
\end{figure}

Once the hyperfine-state dependent loss rates have been determined, they can be used to construct an effective loss rate $\beta^{\prime}$ for use in (\ref{dualEq}).
To find the effective loss coefficients used in (\ref{dualEq}), the state dependent values are weighted by the observed hyperfine state distribution and summed.
The weighing was preformed using our best estimate of the actual hyperfine state distribution incorporating all of our available measurements.
The weighted loss coefficients are the final piece of information required to construct a theoretical dual loading curve for either isotope of rubidium under the independent load rate assumption.
These curves appear with the dual loading data in figure~\ref{Dual}.
The shaded area in figure~\ref{Dual} is due to the inability to separate homonuclear collisions between atoms which are both in the upper hyperfine state vs. between an atom in the upper hyperfine state and an atom in the lower hyperfine state.
This presents a problem when applying the homonuclear loss rate correction to handle the F=2 state population reduction of $^{87}$Rb due to the presence of $^{85}$Rb.
The solid line represents the case where it is assumed that the sole loss mechanism is between atoms in different hyperfine states (F=2+1).
Our previous work~\cite{Hamilton2009} found that the upper hyperfine state fraction was much smaller than the lower hyperfine state fraction which indicates that there are few collisions between two upper hyperfine state atoms compared with collisions between atoms in the upper and lower hyperfine states.
This implies that the behaviour is most likely best modeled by the region close to the solid line.
 
A comparison between the theory model and the observed dual isotope loading behaviour shows a clear and significant difference between the two.  
These differences ultimately caused us to reexamine our independent load rate assumption.
The assumption seemed reasonable because the resonant frequencies of the two isotopes are hundreds of natural linewidths apart, meaning that the atom response to the resonant light of the other isotope is minimal.
The weak atom response to the resonant light of the other isotope was confirmed experimentally by adding the off-resonant light right when the optical trap is turned on so that the off-resonant light is present but the other isotope is not.
Despite this, additional measurements showed that the isotopes could have an effect on each others' load rates despite the large difference in resonant frequencies.
The rest of this article will show why the assumption does not hold.

\begin{figure}
\includegraphics{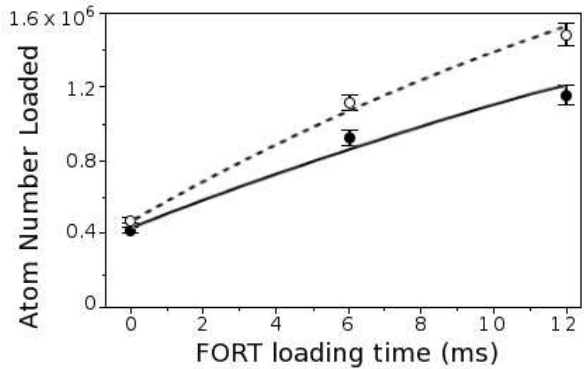}
\caption{\label{Rate_Reduction} FORT loading data for $^{87}$Rb with and without $^{85}$Rb present. 
The plot shows number of $^{87}$Rb atoms in the FORT versus load time. 
Open circles are $^{87}$Rb alone, while full circles are $^{87}$Rb in the presence of an $^{85}$Rb MOT. 
Error bars reflect statistical error of the measurements.
The dashed and solid lines are fits to the $^{87}$Rb alone and $^{87}$Rb in the presence of $^{85}$Rb data respectively.
Only minimal loading of $^{85}$Rb into the FORT was allowed so as not to produce significant light assisted collisional losses.
The impact of these losses can be seen in the reduction of the slope of the loading curve with higher atom number.}
\end{figure}

To observe the cross-isotope hindrance to the load rate, we examined the loading of $^{87}$Rb into the FORT with $^{85}$Rb present.
This was done by first preparing both $^{87}$Rb and $^{85}$Rb MOTs as if doing a dual-isotope experiment.
During the FORT loading sequence, one or both of the $^{85}$Rb lasers were turned off so that $^{85}$Rb was not actively loaded into the FORT.
However, $^{85}$Rb atoms were still present while $^{87}$Rb was loaded into the FORT.
The number of $^{87}$Rb atoms loaded into the FORT as a function of time was then measured and used to determine the load rate (in atoms/s).
The experiment was repeated without having any $^{85}$Rb present by either detuning the $^{85}$Rb MOT lasers to the point that the MOT could not load, or by turning off the trapping or hyperfine pump beams of the $^{85}$Rb MOT.
The load rates of $^{87}$Rb were then compared with and without $^{85}$Rb present during the loading process.
The results of these measurements are shown in figure~\ref{Rate_Reduction}.
The load rates extracted from the data presented in figure~\ref{Rate_Reduction} were 1.129$\pm$0.051x10$^{8}$atoms/sec when loading $^{87}$Rb alone, and 0.821$\pm$0.044x10$^{8}$atoms/sec when loading in the presence of $^{85}$Rb.
This corresponds to a 27$\pm$5$\%$ decrease in the overall loading rate of $^{87}$Rb due to the presence of $^{85}$Rb.
We found that this decrease was not sensitive to the loss rates; variations of 50\% on the values of the loss coefficients yielded no noticeable effect on the proportionate decrease of the load rate.
We note that although $^{85}$Rb was not actively loaded into the FORT, a small number (0.4x10$^{6}$) of $^{85}$Rb ended up in the FORT trapping region during these measurements.
This is partially due to atoms being immediately loaded upon turning on the FORT~\cite{Wu2006}, but also due to non-trapped atoms passing through the trapping region during the experiment contributing to the net density of $^{85}$Rb.

\begin{figure}
\includegraphics{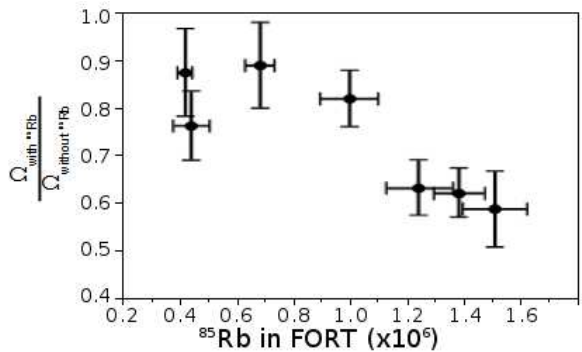}
\caption{\label{Hyper_Dist} Change in the fractional amount of $^{87}$Rb in the F=2 ground state ($\Omega$) as a function of number of $^{85}$Rb loaded into the FORT. 
At small numbers of $^{85}$Rb in the trap, there is small change in the $^{87}$Rb state distribution. 
However, as the number of $^{85}$Rb increases in the FORT, there is significant change to the ground state distribution of the $^{87}$Rb.  
When $^{85}$Rb was absent, the ratio $\Omega_{\rm{with85Rb}}/\Omega_{\rm{without85Rb}}$ was by definition 1 and the value of $\Omega_{\rm{without85Rb}}$ was about 0.25.}
\end{figure}

In addition to observing a decrease in the load rate due to the presence of another isotope, we also found that the presence of both isotopes affected the hyperfine state distributions of the atoms.
For these measurements both isotopes were loaded into the FORT.
$^{85}$Rb was loaded for 0-20~ms prior to being put into the \textit{F}=2 state.
This allowed the $^{85}$Rb atom number loaded into the FORT to be deliberately adjusted.
The $^{87}$Rb was allowed to continue loading before abruptly shutting off both trapping and hyperfine pump MOT beams so as to preserve its hyperfine distribution.
Load time for the $^{87}$Rb was adjusted to preserve the same number of atoms for the duration of the experiments.
After allowing 100~ms for atoms to fall away once loading is completed, the atoms were imaged using standard absorption imaging techniques.
Turning the hyperfine pump beam on or off during imaging allows for the hyperfine state distribution to be determined.
The fraction of $^{87}$Rb in the \textit{F}=2 state ($\Omega$) was then compared for the experiment with and without $^{85}$Rb present.
Figure~\ref{Hyper_Dist} shows the fractional change in $\Omega$ ($\Omega_{with 85Rb}$/$\Omega_{without 85Rb}$) as a function of number of $^{85}$Rb loaded into the FORT.
The data shows that as the number of $^{85}$Rb in the FORT increased, there was a measurable difference in the ground state distribution of $^{87}$Rb.
This strongly implies that the $^{85}$Rb changes the optical pumping of $^{87}$Rb in the FORT, which will affect the optical cooling rate and thus the load rate into the trap.

The observed reduction of load rate and cooling disruption cannot be explained by either reabsorption or typical ultracold collisions (elastic, hyperfine changing, spin-exchange, and light-assisted) for these scattering rates are too low because the associated cross-sections are too small.
However, laser light will induce dipoles in the atoms which can interact, and it turns out that even off-resonant induced dipoles are not inconsequential for our experimental conditions.
Estimates of the induced dipole-dipole forces show that they have a significant influence on the interatomic potential.
For instance, the dipole-dipole interaction strength is large enough that previously closed collision channels (e.g. p-wave, d-wave) become accessible at temperatures present in the gas.
This results in not only an increased elastic collision rate but an increased inelastic, \textit{m}$_F$-state changing collision rate as well.
While not leading to large increases in the atoms' kinetic energy, such collisions can decrease the load rate \textit{R} by disrupting the \textit{m}$_F$ state coherences necessary for effective cooling~\cite{Chang2002}.
A drop in load rate of $35\pm 6\%$ and $37\pm 6\%$ for $^{85}$Rb and $^{87}$Rb respectively accounts for the discrepancy in figure~\ref{Dual}, consistent with the previously-measured load rate reduction after taking into consideration density variations in the MOTs between these measurements.

As a check, we investigated the necessary increase in heteronuclear losses which would remove the discrepancy in figure~\ref{Dual}.
We found that an increase of a factor of 2.5 would be sufficient, but this is well outside the uncertainty of our measurements.
Thus our observations that the interisotope load rate influences must be included for a proper understanding of the dual isotope loading.

To summarize, when loading $^{85}$Rb and $^{87}$Rb from MOTs into an optical trap it is expected that that off-resonant homonuclear and heteronuclear light assisted collisions reduce the maximum number of atoms loaded into the FORT for each isotope as compared to loading the isotopes alone.
We have characterized these loss channels in a simultaneous load of $^{85}$Rb and $^{87}$Rb through explicit measurements.
We find that the sum of the losses due to these additional channels is comparable to the sum of the on-resonant homonuclear losses during single-isotope loading.
However, these additional losses are not enough to explain the observed reduction in the number of atoms which can be loaded into the FORT.
A reduction in the load rate for both $^{85}$Rb and $^{87}$Rb due to the presence of the other isotope can explain the discrepancy in a manner which is consistent with additional observations.
Our results thus indicate the significance of both light-assisted collisional losses and laser cooling efficiency disruption in the performance of loading $^{85}$Rb and $^{87}$Rb into a shallow FORT.
It is expected that similar effects would be present in other experiments which load dual atom species into a FORT.

\ack
This work is funded by the Air Force Office of Scientific Research, Grant No. FA9550-08-1-0031.

\bibliographystyle{unsrt_mod_iop}

\end{document}